\documentclass[letterpaper, 10 pt, conference]{ieeeconf}  % Comment this line out if you need a4paper

\IEEEoverridecommandlockouts                              % This command is only needed if 
\makeatletter
\long\def\@makecaption#1#2{\ifx\@captype\@IEEEtablestring%
	\footnotesize\begin{center}{\normalfont\footnotesize #1}\\
		{\normalfont\footnotesize\scshape #2}\end{center}%
	\@IEEEtablecaptionsepspace
	\else
	\@IEEEfigurecaptionsepspace
	\setbox\@tempboxa\hbox{\normalfont\footnotesize {#1.}~~ #2}%
	\ifdim \wd\@tempboxa >\hsize%
	\setbox\@tempboxa\hbox{\normalfont\footnotesize {#1.}~~ }%
	\parbox[t]{\hsize}{\normalfont\footnotesize \noindent\unhbox\@tempboxa#2}%
	\else
	\hbox to\hsize{\normalfont\footnotesize\hfil\box\@tempboxa\hfil}\fi\fi}
\makeatother
\usepackage{graphicx}
\usepackage{float}
\usepackage{ctable}
\usepackage{algorithm}
\usepackage{wrapfig}
\usepackage{algpseudocode}
\usepackage{multirow}
\usepackage{amsfonts}
\usepackage{mathrsfs}
\usepackage{amsmath}
\graphicspath{{figs/}}

\usepackage[noadjust]{cite}

\usepackage[T1]{fontenc} % optional enhanced font encoding
\usepackage{amsmath}
\usepackage[cmintegrals]{newtxmath}

\usepackage{bm}

\usepackage{array}

\usepackage{url}

\providecommand{\hypersetup}[1]{\relax}

\hypersetup{pdftitle={Bare Demo of IEEE\_lsens.cls for IEEE Sensors Letters},%<!CHANGE!
pdfsubject={Typesetting},%<!CHANGE!
pdfauthor={Michael D. Shell},%<!CHANGE!
pdfkeywords={Class, IEEE, IEEE\_lsens, IEEE Sensors Letters, LaTeX, Typesetting, TeX}}%<^!CHANGE!}

\usepackage{color}
\newcommand{\vect}[1]{\boldsymbol{#1}}
\graphicspath{{figs/}}
\begin{document}

\title{\LARGE Physics-constrained Active Learning for Soil Moisture Estimation and Optimal Sensor Placement}

\author{Jianxin Xie$^*$, Bing Yao, Zheyu Jiang
\thanks{
	Jianxin Xie ($^*$corresponding author: hcf7fd@virginia.edu) is with the School of Data Science, University of Virginia, Charlottesville, VA 22904 USA; Bing Yao is with the Department of Industrial and Systems Engineering, University of Tennessee at Knoxville, TN 37996 USA; Zheyu Jiang is with the School of Chemical Engineering, Oklahoma State University, Stillwater, OK 74078 USA.
}}

\maketitle

\begin{abstract}
		
Soil moisture is a crucial hydrological state variable that has significant importance to the global environment and agriculture. Precise monitoring of soil moisture in crop fields is critical to reducing agricultural drought and improving crop yield. In-situ soil moisture sensors, which are buried at pre-determined depths and distributed across the field, are promising solutions for monitoring soil moisture. However, high-density sensor deployment is neither economically feasible nor practical. Thus, to achieve a higher spatial resolution of soil moisture dynamics using a limited number of sensors, we integrate a physics-based agro-hydrological model based on Richards’ equation in a physics-constrained deep learning framework to accurately predict soil moisture dynamics in the soil’s root zone. This approach ensures that soil moisture estimates align well with sensor observations while obeying physical laws at the same time. Furthermore, to strategically identify the locations for sensor placement, we introduce a novel active learning framework that combines space-filling design and physics residual-based sampling to maximize data acquisition potential with limited sensors. Our numerical results demonstrate that integrating Physics-constrained Deep Learning (P-DL) with an active learning strategy within a unified framework—named the Physics-constrained Active Learning (P-DAL) framework—significantly improves the predictive accuracy and effectiveness of field-scale soil moisture monitoring using in-situ sensors.
		
\end{abstract}

%\begin{IEEEkeywords}
%Soil moisture, Richards equation, physics-informed deep learning, sensor placement, active sampling.
	
%\end{IEEEkeywords}

\section{Introduction}
    Soil moisture is a key hydrological state variable that has significant importance for the global environment and human society~\cite{robinson2008soil}. In particular, accurate modeling and monitoring of root zone soil moisture in crop fields, which defines the amount of water stored within the plant root zone (top 100 cm of soil) available for transpiration and photosynthesis, is essential for improving agricultural production and crop productivity, providing a basis for precision irrigation and agriculture, preventing leaching of agrochemicals and soil nutrients into groundwater, and predicting agricultural droughts~\cite{babaeian2019ground}.
    
    Physics-based models, formulated as partial differential equations (PDEs), %that characterize the water flow dynamics in a porous medium such as soil, 
    have been developed to quantitatively understand the transport behavior of root-zone soil moisture. %and assess the impact of soil moisture on the environment and agricultural activities. 
    These PDEs can be solved using various numerical methods to extract exact solutions \cite{ogden2017soil}. For example, algorithms that rely on mesh structures, such as the Finite Element Method (FEM), are extensively employed for the simulation and visualization of soil moisture dynamics \cite{vsimuunek2006hydrus}. %farthing2017numerical
    Recently, progress was made in combining finite volume discretization and neural networks to improve the accuracy of mesh-based numerical schemes \cite{song2023escape, song2023data}. Despite these recent advancements, the solution quality of mesh-based approaches typically depends on spatial and temporal discretization.
   %The computation complexity to solve the RE at each time step is proportional to the number of discretized nodes within the targeted crop field, resulting in expensive computational costs for detailed modeling. More importantly, real-world sensor measurements cannot be readily assimilated into the FEM numerical procedure, leading to inferior applicability of mesh-based simulation in real-world practice.
   Specifically, the computational effort needed to solve the discretized PDEs at each time step increases with the number of discretized nodes, leading to high computational costs for in-depth modeling. Furthermore, the numerical process to solve these physics-based models does not readily incorporate the actual sensor data, which limits the accuracy and practical usage of pure physics-based models in real-world applications.

   Using soil moisture sensor observations, traditional machine learning models such as decision tree, support vector machine, and $k$-nearest neighbor have been successfully applied to address soil moisture problems \cite{pasolli2011estimating, liu2017comparison, liu2020generating}. %pekel2020estimation, 
   However, it is reported that these traditional approaches present weak robustness and tend to generate unstable predictions \cite{liu2020generating, ali2015review}. Recently, with the rapid development in artificial intelligence, deep learning methods have achieved high predictive power and strong fitting capability to nonlinear, non-explicit functional relationships \cite{xie2022physics, shen2017deep,%srivastava2021comprehensive,
   	 cai2019research}. Compared to traditional regression approaches, deep learning is more capable of processing big data for better predictive performance \cite{lecun2015deep}. Deep learning has a wide spectrum of applications in soil and water-related applications and can better capture the complex spatiotemporal dynamics of soil moisture \cite{ali2015review}. For example, Cai et al. \cite{cai2019research} constructed a deep-learning regression network to predict soil moisture using features extracted from the Taylor diagram. %Padarian et al. \cite{padarian2019using} developed a convolutional neural network to predict soil properties from raw soil spectra that contains soil physicochemical and biological properties. Song et al. \cite{song2016modeling} engaged a macroscopic cellular automata model by deep belief network to predict the soil moisture content over an irrigated corn field. Lee et al. \cite{lee2019estimation} implemented a deep learning method for reliable estimation of soil moisture based on remotely sensed satellite data.  
   Yu et al. \cite{yu2021hybrid} developed a hybrid convolutional neural network-gated recurrent architecture to predict soil moisture in a mazie root zone given the soil water content and meteorological observation. Li et al. \cite{li2022causality} proposed a causality-structure-based Long Short-Term Memory (LSTM) network with enhanced model interpretation of time interdependency and causality to predict surface soil moisture.  
   
   A practical challenge for accurately measuring soil moisture profile using in situ sensors arises from the fact that it is unaffordable, tedious, and environmentally destructive to deploy in situ soil moisture sensors everywhere in a field. This forces sensor observations to be made at a coarse level and leads to the following question: ``How can practitioners strategically place a limited number of in situ soil moisture sensors in a field while achieving the best root-zone soil moisture estimation for the whole field?'' To address this question, we propose a physic-constrained deep active learning (P-DAL) framework. In particular, our contributions are summarized as follows:
  \begin{itemize}
  	\item[1.] We incorporate a physics-based PDE model that governs water flow dynamics in soil into a deep learning framework to inform the prediction. The resulting soil moisture predictions obey both the sensor observations and the governing transport phenomena. 
  	\item[2.] We propose a novel active learning framework to guide the sequential optimal sensor placement in a field, such that the selected sensor locations would provide the most information needed for accurate soil moisture estimation.
  	\item [3.] We systematically validate the effectiveness of our proposed P-DAL framework by conducting simulation experiments for both evaporation and infiltration scenarios.
  \end{itemize}

\section{Related Prior Work}
    \subsection{Physics-informed deep learning for root-zone soil moisture estimation}
    
    Physics-informed machine learning (PIML) is a powerful tool that incorporates the prior knowledge of physical laws and the actual sensor observations in a data-driven framework. As a result, PIML can overcome the low data availability issue that would limit the capability of most machine learning models. PIML incorporates known physical laws as constraints during training, enhancing its ability to generalize beyond data, improve interpretability, and guide predictions in accordance with underlying physical laws. For example, Raissi et al. \cite{raissi2019physics} built a physics-informed neural network (PINN) framework that integrates well-established physics laws with deep learning to suppress the model dependence on training data. The efficacy of PINN has already been verified in numerous physical systems, such as the fluid dynamics \cite{raissi2020hidden }, solid mechanics \cite{rao2021physics, zhang2020physics}, heat transfer \cite{cai2021physics}, and biological systems \cite{xie2022physics, xie2022physics1}.%,sahli2020physics}. 
    
    In terms of root-zone soil moisture estimation, most existing agro-hydrological models are based on the Richards equation (RE) \cite{richards1931capillary}, which captures irrigation, precipitation, evapotranspiration, runoff, and drainage dynamics in soil. Recently, researchers have investigated the application of PINNs to model soil moisture dynamics by incorporating the RE. Notably, Tartakovsky et al. \cite{tartakovsky2018learning} were one of the first teams to utilize PINNs to derive the hydraulic conductivity function in unsaturated homogeneous soil using pressure head data based on the 2D RE. Banbai et al. \cite{bandai2021physics} embedded RE into PINN to inversely learn the soil moisture dynamics only from soil sensor measurements without engaging any pre-assumptions on soil hydraulic functions and realize a free-form representation of constitutive relationships. Depina \cite{depina2022application} inherited and extended Banbai et al.'s work and utilized PINN to investigate the 1D solutions of the RE that adopts the van Genuchten constitutive model, which allows a simpler neural network structure for pressure head estimation. More recently, Haruzi et al. \cite{haruzi2023modeling} proposed a PINN model with non-invasive geometric data to simulate 2D water flow and solute transport. These studies mainly focus on the application of PINN in 1D or 2D soil systems. On the other hand, this work generalizes the predictive capabilities of PINNs to 3D soil systems. Meanwhile, integrating underlying physics (i.e., the RE) into deep neural networks can help reduce reliance on extensive sensor measurements, however, the performance of these predictive models still significantly depends on the volume and quality of training data \cite{xie2022physics1}. 
    
    \subsection{Optimal soil moisture sensor placement}
    
    As mentioned earlier, it is unaffordable and impractical to deploy sensors everywhere in the field. Conventional grid or random sampling strategies based on heuristics (ranging from 2 sensors/100 acres \cite{numberheuristics} to 20 sensors/acre \cite{uflsensor}) are also arbitrary and ineffective. For in situ soil moisture sensing, more systematic sensor placement algorithms have been developed to better infer field-wide soil moisture profile from sparse sensor measurements. 
    Wu et al. \cite{wu2012situ} used statistical clustering with the Gaussian process to find a coarse-grained monotonic ordering of locations in terms of the soil moisture content. Specifically, they classified the clusters based on the order of the mean and the number of sensors allocated to each cluster is decided based on the variance's magnitude. %by the order of the variance. 
    Dursun et al. \cite{dursun2017optimization} developed a generic algorithm that iteratively refines soil moisture sensor locations. This algorithm works by continuously eliminating the least effective sensor position and replacing it with the most optimal candidate from the current iteration.
    Sahoo et al. \cite{sahoo2019optimal} propose to estimate the soil moisture dynamics in agro-hydrological systems with the Kalman filter. %They first reduce the order of RE to reduce the computational complexity.
    They use the graphic approach with structural observability to identify the minimal number of sensors, followed by using the modal degree of observability to find their optimal placement. 
    However, these optimal sensor placement approaches stem from either a statistical perspective or a graphical understanding. They did not consider the underlying physics rules to guide the search, which makes the selection devoid of fundamental physics insights. 
    %This necessitates an effective active learning strategy that guides the selection of sensor locations, ensuring they capture comprehensive and informative data about soil moisture variations.
   % it has never been considered to identify the sensor placement for soil moisture given physics-informed predictive models. 

\section{Research Methodology}

\subsection{The Richards Equation}
The soil moisture dynamics are fundamentally governed by the Richards equation (RE) \cite{richards1931capillary}. Without loss of generality of our P-DAL framework, we consider the scenarios where the sink term accounting for root water uptake is negligible. The resulting continuity equation that models the mass balance of water in a soil system is written as:
%In this paper, we engage non-linear RE to discribe the flow of water in 3D unsaturated d soil and ingore the sink term: % soil moisture hysteresis
\begin{equation}
	\label{Eq:RE}
	\frac{\partial \theta (\psi)}{ \partial t} = -\nabla\cdot q
\end{equation}
where $\theta$ is the volumetric water content in the soil (i.e., soil moisture), $\psi $ stands for the pressure head, $t$ denotes time, and $q$ represents the water flux. In addition to the continuity equation, the RE incorporates the Buckingham-Darcy law \cite{buckingham1907studies}, which extends the traditional Darcy's Law to account for the capillary forces in unsaturated soils. This is characterized by the relationship between $q$ and $\psi$: 
\begin{equation}
	\label{Eq:Darcy}
	q = -K(\psi)\cdot\nabla (\psi +z)
\end{equation}
where $K$ is the hydraulic conductivity. By incorporating Buckingham-Darcy's law of Eq. \eqref{Eq:Darcy} into Eq. (\ref{Eq:RE}), the RE can be expressed as:
\begin{equation}
	\frac{\partial \theta (\psi)}{\partial t}=\nabla \cdot(K(\psi) \nabla(\psi+z))
	\label{Eq:re}
\end{equation}
 
It is worth noting that the pressure head $\psi$ is a spatiotemporal variable linked to both time $t$ and spatial coordinates $\vect{s} = [x,y,z]$. Thus, the left-hand side of Eq. (\ref{Eq:re}) can be explicitly written as $\frac{\partial \theta}{\partial \psi} \frac{\partial \psi}{\partial t}$ by the chain rule. 

Both the hydraulic conductivity $K$ and soil moisture $\theta$ are highly nonlinear functions of pressure head $h$ and soil properties. Specifically, $\theta(\psi)$ and $K(\psi)$ are commonly referred to as the water retention curve (WRC) and hydraulic conductivity function (HCF), respectively. Both WRC and HCF have been regressed and tabulated as parametric models for various soil types \cite{assouline2006modeling, brooks1964hydraulic}. %kosugi1994three , iden2014comment
Without loss of generality, in this study, we adopt the widely used van Genuchten model \cite{van1980closed} for both WRCs and HCFs:

\begin{equation}
	\begin{aligned}
		\theta (\psi) &= \frac{\theta_s - \theta_r}{\left[1 + (\alpha 	|\psi |)^n \right]^m}+\theta_r,\\
		%	C(\psi) = \frac{d\theta}{d\psi} = \frac{mn\psi (-(\theta_s-\theta_r))\alpha^n |\psi|^{n-2}}{(\alpha^n |\psi|^n +1)^{m+1}} \\
		K(\psi) &= K_s\frac{ \big\{ 1 - (\alpha |\psi|)^{n-1} [1 +  (\alpha |\psi|)^n]^{-m}  \big\}^2}{[1+(\alpha |\psi|)^n]^{m/2}},
	\end{aligned}
	\label{Eq:con}
\end{equation}
where $K_s, \theta_s, \theta_r$ represent the saturated hydraulic conductivity, saturated volumetric moisture content, and residual moisture content, respectively. Parameters $n$, $m=1-1/n$, and $\alpha$ stand for curve-fitting soil hydraulic properties. The values of these parameters are taken from \cite{celia1992mass} for this study. %These soil hydraulic parameters are determined by the soil properties of the crop field \cite{orouskhani2022impact}, the values of which are taken from \cite{celia1992mass}.

We examine a 3D cuboid soil represented in a $xyz$ Cartesian coordinate system. In our study, we consider two scenarios, namely evaporation and infiltration, that model the moisture leaving and entering the soil surface, respectively. For evaporation, we adopt the Neumann boundary condition in the RE for all 6 faces (i.e., north, south, west, east, top, and bottom) of the cuboid soil geometry as follows:
\begin{align}
	\nabla{\psi}(\vect{s},t) &= 0 \qquad \text{if} \quad x=0, L \text{ or } y=0,W \text{ or } z=0  \nonumber\\
	\nabla{\psi}(\vect{s},t) - c_1 &=0\qquad \text{if} \quad z = D \nonumber \\
	\label{Eq:boundary1}
\end{align}
where $L, W, D$ denotes the length, width, and depth of the soil cuboid. When analyzing infiltration in the presence of rainfall, we adopt the Neumann boundary condition for the vertical boundaries (i.e., the north, south, west, and east faces), and the Dirichlet condition for the top and bottom surfaces:
\begin{align}
	\nabla{\psi}(\vect{s},t) &= 0 \qquad \text{if} \quad x=0, L \text{ or } y=0,W \nonumber\\
	{\psi}(\vect{s},t) -c_2 &= 0 \qquad \text{if} \quad z = 0 \nonumber \\
    {\psi}(\vect{s},t) - c_3&= 0 \qquad \text{if} \quad z = D \nonumber \\
    \label{Eq:boundary2}
\end{align}

Note that $c_1, c_2, c_3$ are constants. These boundary conditions characterize the behavior of pressure head $\psi$ at the boundary of the 3D land.

%\subsection{Physics-constrained deep active learning (P-DAL)}

\subsection{Physics-constrained deep learning (P-DL) framework}
\begin{figure*}
	\centering
	\includegraphics[width=5.5in]{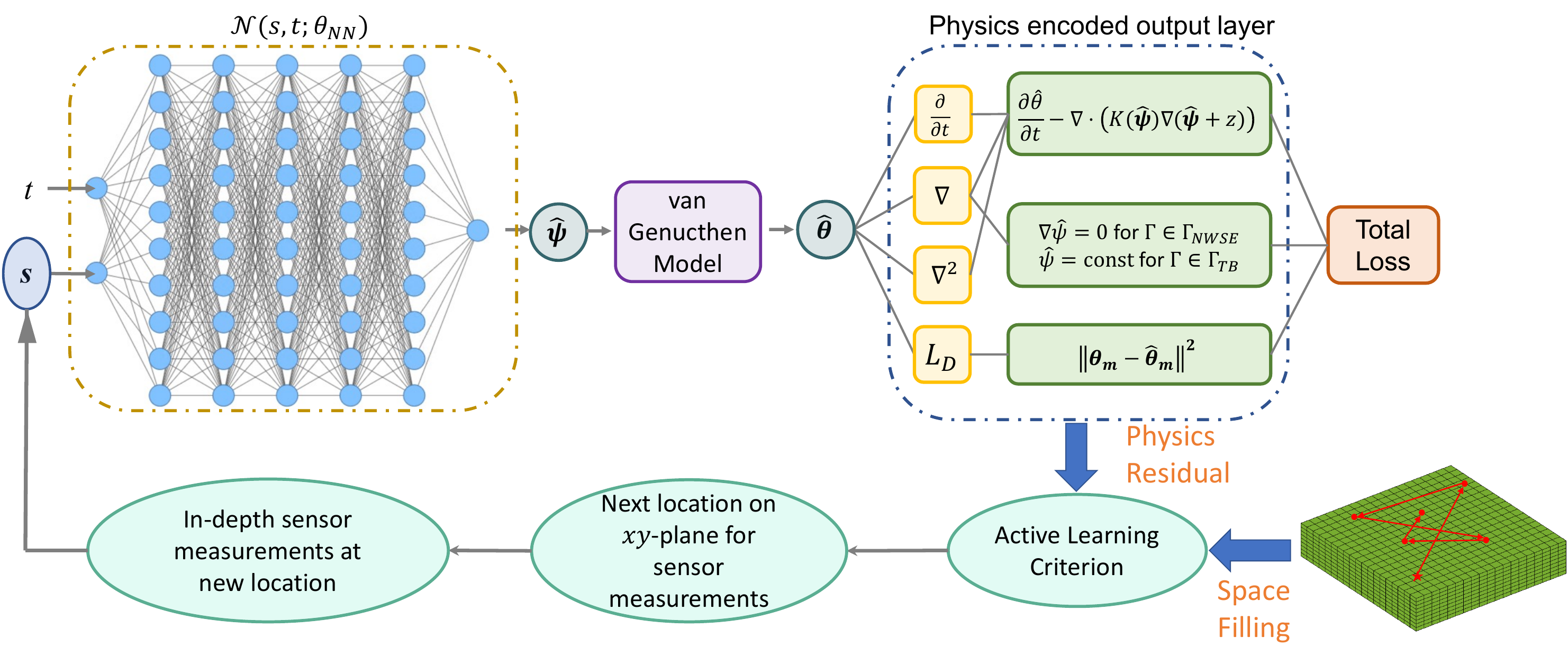}
%\centerline{\includegraphics{fig1.png}}
    \vspace{-15pt}
	\caption{Illustration of the proposed P-DAL framework for soil moisture prediction. The RE and the boundary conditions are transformed into residuals as the physics-based loss, which is incorporated into the total loss function. $\Gamma_{NWSE}$ stands for the north, west, south, and east vertical faces. And $\Gamma_{TB}$ denotes the top and bottom horizontal faces of the soil geometry.}
	\vspace{-15pt}
	\label{Fig:flowchart}
\end{figure*}

Fig. \ref{Fig:flowchart}  illustrates our proposed Physics-constrained Deep Active Learning (P-DAL) framework. This framework engages a physics-constrained neural network (P-DL) \cite{xie2022physics, xie2022physics1} as a cornerstone to predict the spatial and temporal variations in soil moisture with in situ sensor observations of soil moisture content. Building on the P-DL model, we develop an innovative active learning scheme to identify the most informative locations for placing subsequent soil moisture sensors, thereby enhancing soil moisture prediction of the entire land produced. This active learning strategy employs a combination of physics-informed residual-based sampling and a space-filling design across the land, which will be elaborated in Section \ref{Sec:active}. % This active learning approach dynamically  By doing so, it can significantly enhance the predictive capability of the overall P-DL model, thereby ensuring more accurate and reliable soil moisture characterization of the entire land.

The characterization of soil moisture majorly depends on the accurate modeling of $\psi(\vect{s},t)$ and $\theta(\vect{s},t)$. We achieve this by using a fully connected feedforward deep neural network (DNN) to approximate the nonlinear relationships between the input spatiotemporal instances $(\vect{s}, t)$ and the distribution of the pressure head $\psi$. The DNN output, denoted as $\hat{\psi}$,  is anticipated to fulfill two primary conditions: firstly, it should align with the sensor measurements of volumetric moisture content $\theta_m$, as depicted by the WRC function in Eq. \ref{Eq:con}; and secondly, it must adhere to the fundamental physical reality, i.e., RE. Specifically, we model the spatiotemporal pressure head distribution as:
$$[\vect{s},t] \xrightarrow{\mathcal{N}\left(s, t ; \Theta_{N N}\right)} \hat{\psi}(\vect{s}, t)$$
where $\mathcal{N}\left(s, t ;\Theta_{N N}\right)$ is the DNN and $\Theta_{NN}$ denotes the DNN parameters. The DNN  contains an input layer encompassing space-time instances $[\vect{s}, t]$, several hidden layers to approximate functional relationships between the input and output, and one output layer to estimate $\hat{\psi}(s,t,\Theta_{NN})$. The RE is further embedded into the DNN, together with in situ sensor observations, to form a new loss function defined as:
\begin{equation}
	\mathcal{L}(\Theta_{NN}) = \mathcal{L}_D+ \mathcal{L}_{Phy}
	\label{Eq:pdl}
\end{equation}

The total loss $\mathcal{L}(\Theta_{NN}) $ consists of the following two components:

1) \textbf{Data-driven loss} $\mathcal{L}_{D}$: The soil moisture content is measured at multiple locations on the horizontal plane of the field (the $xy$-plane), as well as at differing depths (the $z$ direction) for each selected 2D location. Every sensor captures a time series of soil moisture signals represented by $ \theta_m(\vect{s},t) $. The DNN is trained to produce predictions, $\hat{\psi}$, that align closely with the actual soil moisture sensor readings, i.e., $\theta_m(\vect{s}, t)$. Recall that the predicted pressure head $\hat{\psi}$ is related to $\theta$ through the WRC function (i.e., Eq. (\ref{Eq:con})). Hence, the data-driven loss $\mathcal{L}_{D}$, enforcing agreement between the sensor observations and estimated pressure head $\hat{\psi}_m$ values at the placement locations, is formulated as:
\begin{equation}
	\mathcal{L}_{D}=\frac{1}{N_m}\sum_{i=1}^{N_m}(\theta_m(\vect{s}_i, t_i)-\theta(\hat{\psi}_m(\vect{s}_i, t_i)))^2
	\label{Eq:loss_hb}
\end{equation}
where ${N_m}$ is the total number of spatiotemporal measurements.
%In order to reduce the dependence of the labeled data that is invasively collected from heart surface, additional physics constraints are engaged to regularize the DNN-based infrastructure for the robust and precise HSP mapping reconstruction. 

(2) \textbf{Physcis-based loss} $\mathcal{L}_{Phy}$: To improve the predictive accuracy and robustness of the DNN model, we introduce a physics-based constraint at the spatiotemporal collocation points $[\vect{s}_i,t_i], i=1,...,N_c$, where $N_c$ is the total number of collocation points. These points are randomly selected from the spatiotemporal domain in the target land to encode the physics knowledge for reinforcing the prediction's adherence to the RE, i.e., Eq. \eqref{Eq:re}. Specifically, the RE-based residual is defined as:
\begin{eqnarray}
	r(s,t,\Theta_{NN}) := \frac{\partial \theta (\hat{\psi})}{\partial t}-\nabla \cdot(K(\hat{\psi})\nabla(\hat{\psi}+z))
	\label{Eq:res}
\end{eqnarray}
%It is worth noting that the gradient and partial derivative in Eq.\ref{Eq:res} can be reformulated in terms of $\psi$ and its derivatives:
%\begin{align}
%	r(s,t,\Theta_{NN}) := \frac{\partial \theta}{\partial \psi} \frac{\partial \psi}{\partial t} &- \left(\frac{\partial K}{\partial \psi} \frac{\partial \psi}{\partial x} \frac{\partial \psi}{\partial x}+K\frac{\partial^2\psi}{\partial x^2}  \right. \nonumber \\
%	&\left. + \frac{\partial K}{\partial \psi} \frac{\partial \psi}{\partial y} \frac{\partial \psi}{\partial y}+K\frac{\partial^2\psi}{\partial y^2} \right. \nonumber \\
%	&\left. + \frac{\partial K}{\partial \psi} \frac{\partial \psi}{\partial z}\left( \frac{\partial \psi}{\partial z}+1\right)+K\frac{\partial^2\psi}{\partial z^2} \right)
%\end{align}
%The gradient $ \frac{\partial \theta}{\partial \psi}$ and $ \frac{\partial K}{\partial \psi}$ can be directly calcualted from the explicit function relationships in Eq. \ref{Eq:con}. 
The first and second-order partial derivatives of $\psi$ can be efficiently calculated through automatic differentiation, a technique developed for backpropagation in deep learning \cite{paszke2017automatic}. The physics-based constraint is enforced by optimizing $r_{\psi}(\vect{s},t;\Theta_{NN})$ towards zero. Consequently, the RE-based loss is defined as:
\begin{equation}
	\mathcal{L}_{RE}=\frac{1}{N_f}\sum_{i=1}^{N_{f}}\|r(s_i,t_i;\Theta_{NN})\|^2
\end{equation}
where $N_f$ is the total number of selected collocation points to enforce the RE.

 Similarly, the boundary conditions are incorporated in the model in terms of the boundary-related residuals. The boundary conditions in Eq. (\ref{Eq:boundary1}-\ref{Eq:boundary2}) can be concisely represented as $\mathcal{B}(\psi, \vect{s}, t)=0$ on $\Gamma$ which stands for the boundaries. To ensure that the prediction $\hat{\psi}$ is consistent with the boundary conditions, we define the boundary condition-based loss as:
\begin{equation}
\mathcal{L}_\mathcal{B} = \frac{1}{|\Gamma|}\sum_{\vect{s}\in \Gamma}\|\mathcal{B}(\hat{\psi},\vect{s},t)\|^2_2
\end{equation}

 Then, $\mathcal{L}_\mathcal{B}$ joins the RE-based loss to create the physics-based loss: $\mathcal{L}_{Phy} = \mathcal{L}_{RE} + \mathcal{L}_{\mathcal{B}} $, which is then combined with the data-driven loss (Eq. (\ref{Eq:loss_hb})) to formulate the overall loss function in Eq. (\ref{Eq:pdl}). This loss setting in DNN allows for a comprehensive consideration of both sensor readings and the fundamental physics governing the water flow dynamics in soil, which will enable the reliable modeling of the spatiotemporal soil moisture dynamics.

%%%%%%%%%%%%%%%%%%%%%%%%%%%%%%%%%%%%%%%%%%%%%%%%%%%%%%%%%%%%%%%%%%%%%%%%%%%%%%%%%%%%%%%%%%%%%%%%
\subsection{Active learning for optimal sensor placement}
\label{Sec:active}

Even though the involvement of physics regularization can alleviate the model reliance on the training data, the quality and volume of training data can still significantly impact the model performance \cite{lu2021deepxde,xie2022physics1}. The cost associated with deploying sensors becomes a significant factor for soil moisture monitoring in a large field. Optimal sensor placement is urgently needed to enable the use of a limited number of in situ sensors while still maintaining high-quality predictive modeling of soil moisture dynamics. Note that the high-resolution 3D mapping of soil moisture is predicted in light of the recorded time-series data from sensor placement locations. By strategically positioning sensors, we can capture the spatial variability in soil moisture more accurately, which is vital for quantitatively monitoring soil moisture distribution and managing crop irrigation.
Here, we introduce a novel active learning approach that fuses residual-based sampling with a space-filling strategy. The goal is to collect the most essential time series data for training the P-DL model so that the model outcome remains robust even with a limited number of sensors.
%This is primarily due to the sheer number of sensors required to adequately cover such large areas, the costs of installation, and ongoing maintenance, all of which can accumulate considerably in large-scale operations. 

(1) {\textbf{Residual-based sampling}}:	
%While deep learning is widely recognized for its excellent predictive performance in nonlinear systems, standard DNNs lack the capability to provide information about optimal sensor placement. 
Traditional methods for seeking the sensor location in soil moisture systems mostly depend on exploring the statistical insights of the model output or graphical interrelationships  \cite{wu2012situ,dursun2017optimization,sahoo2019optimal}. These methods ignore the underlying physics truth that governs the soil moisture dynamics. Additionally, due to their specially designed model infrastructure for estimating soil moisture, these sensor placement algorithms may not be readily applied in a deep learning framework. Inspired by the work of Katharopoulos and Fleuret \cite{katharopoulos2017biased}%katharopoulos2018not}
, which demonstrate that the selection of training samples based on loss magnitude can expedite the convergence of neural network optimization, we propose an innovative residual-based sampling strategy for robust prediction of soil moisture using P-DL.

%Katharopoulos and Fleuret \cite{katharopoulos2017biased,katharopoulos2018not} has theoretically demonstrate that the selection of training samples based on loss magnitude can expedite the convergence of neural network optimization. Based on this, an active sampling strategy \cite{nabian2021efficient} is proposed to collect the training points according to the distribution computed from the residual loss. However, this method   This strategy is specifically designed for P-DL and is aimed at optimizing soil moisture sensor placement. 

The proposed residual-based sampling scheme aims to find the most informative location on a horizontal land (i.e., $xy$-plane) by identifying a spatial location with the largest residual value on the 2D plane. Similar to Eq. (\ref{Eq:res}), the residual for every spatial node in the soil geometry and temporal instance can be calculated by:
\begin{align}
	&r(\vect{s}_i,t_j,\Theta_{NN}) := \frac{\partial \theta}{\partial t} - \nabla \cdot \left(K(\hat{\psi}(\vect{s}_i,t_j))\nabla(\hat{\psi}(\vect{s}_i,t_j) + z)\right)\\
&\text{for } i = 1, \dots, N \text{ and } j = 1, \dots, T. \label{Eq:res1} \nonumber
\end{align}
where $\vect{s}_i \in \mathbb{R}^3$, $N$ is the total number of discretized spatial nodes and $T$ is the total number of temporal instances. Let $N_L, N_W, N_D$ be the number of the discretized spatial nodes for the soil geometry's length, width, and depth, respectively. This leads to $N =N_LN_WN_D$. We denote $r_{\vect{\kappa}} $ as the cumulative residual over different depths and time instances for a given location on the $xy$-plane. %It is defined as the sum of residuals $r$ evaluated at each depth $z_d$ and time $t_j$, for $d = 1,...,N_D$ and $j=1,...,T$. The mathemetical formulation can be expressed as:
\begin{align}
	r_{	\vect{\kappa}}(x_l, y_w,\Theta_{NN}) = \sum_{j=1}^T\sum_{d=1}^{N_D} \| r(x_l, y_w, z_d, t_j, \Theta_{NN})\|^2
\end{align}
where $l=1,...,L$ and $w=1,...,W$.
We use min-max normalization to rescales $r_{	\vect{\kappa}}$ to the range $[0,1]$:
\begin{equation}
	r^\prime_{	\vect{\kappa}} = \frac{r_{\vect{\kappa}} - r_{\vect{\kappa}\min}}{  r_{\vect{\kappa}\max} -   r_{\vect{\kappa}\min}}
\end{equation}
where $r^\prime_{\vect{\kappa}}$ represents the normalized values for the residuals in $xy$-plane. $ r_{\vect{\kappa}\min}$ and  $r_{\vect{\kappa}\min}$ stand for the minimum and maximum value of $r_{\vect{\kappa}}$, respectively. The location index exhibiting the largest $r_{\vect{\kappa}}^\prime$ value indicates that the prediction at this specific 2D location deviates most significantly from the established physics-based model, the Richards equation.

The residual-based active sampling may enable faster convergence for neural network training. Lu et al. \cite{lu2021deepxde} first proposed a residual-based adaptive refinement to improve the distribution of residual points during the training process. Based on this, Yu et al. \cite{yu2022gradient} and Wu et al . \cite{wu2023comprehensive} propose to adaptively add training data where the residuals are large to improve the prediction. %{\color{blue}
However, in cases where residuals are non-uniform or have significant variations across the 2D domain, residual-based sampling alone might struggle to adequately identify proper sensor locations that carry global information about soil moisture dynamics. %The sole criterion may also fail if the locations with high residual aggregate together at local regions. (\color{red}The two sentences are confusing and seem contradictory to each other).} These factors may adversely affect the accuracy of the overall solution. 

(2) {\textbf{Space-filling design}}: 
To enhance global soil moisture prediction with P-DL, we propose to further incorporate the maximin-distance design, a space-filling approach for optimizing computer experiments, into our active learning framework. Let $\vect{\kappa} = [x,y]$ denote the spatial location in $xy$-plane. In the conventional sequential learning process that relies on a purely space-filling design, the subsequent query point $\vect{\kappa}_{n+1}$ is determined by:
\begin{eqnarray}
	\vect{\kappa}_{n+1}=\arg\max_{\vect{\kappa}}\min_{i\in\{1,2,...,n\}}\mathrm{dist}(\vect{\kappa},	\vect{\kappa}_i)
	\label{Eq: SF}
\end{eqnarray}
This approach generates the subsequent point $\vect{\kappa}_{n+1}$ by ensuring it has the maximum possible minimum distance from the already observed locations $\vect{\kappa}_i$'s, $i\in\{1,\dots,n\}$.
%which produces the next point $\vect{\kappa}_{n+1}$ that maximizes the minimum distance between $\vect{\kappa}_{n+1}$ and existing observation locations $\vect{\kappa}_i$'s, $i\in\{1,\dots,n\}$. 
The Euclidean distance function $dist(\cdot)$ is employed in Eq. (\ref{Eq: SF}) to account for the spatial interplays in a 2D field. This will then be combined with the residual-based sampling scheme to form a new active learning criterion.

(3) {\textbf{Active learning criterion}}: %\label{s:methods.1.1}
A good active learning (AL) criterion for a physical system shall account for both the deviation degree from the fundamental laws and the distribution level of chosen observation locations. To meet this objective, we design a new AL criterion that integrates the residual magnitude with the max-min design as:
\begin{align}
	\vect{\kappa}_{n+1}=\operatorname{argmax}_{\vect{\kappa}}\left\{   r^\prime_{\vect{\kappa}}(\vect{\kappa} ) +\lambda \cdot \frac{\min\limits _{i \in\{1, \cdots, n\}} \mathrm{dist}\left(\vect{\kappa} -\vect{\kappa} _{i}\right)}{\max \limits_{\vect{\kappa} } \min \limits_{i \in\{1, \cdots, n\}}  \mathrm{dist}\left(\vect{\kappa} -\vect{\kappa} _{i}\right)}\right\}
	\label{Eq: AL}
\end{align}
where  $ \min _{i \in\{1, \cdots, n\}} \mathrm{dist}\left(\vect{\kappa} -\vect{\kappa} _{i}\right) $ is the shortest distance from the unobserved location  $ \vect{\kappa} $ to any of the measured location $\vect{\kappa} _i  $'s, $i\in\{1,\dots,n\}$ on the horizontal plane. The greatest possible value of these minimum distances is expressed as $ \max_{\vect{\kappa} } \min_{i \in\{1, \cdots, n\}}  \mathrm{dist}\left(\vect{\kappa} -\vect{\kappa} _{i}\right) $, which serves to normalize the space-filling criterion. Parameter $ \lambda > 0$ is introduced to balance between the influences of the selection based on residuals and space-filling design. Its value is empirically set as 1 in later numerical experiments. The proposed AL criterion in Eq. (\ref{Eq: AL}) is designed to find the potential sensor locations that carry the most comprehensive information about the entire soil moisture dynamics. The 2D locations indicated by the AL criterion highlight areas with low physical fidelity while simultaneously considering the global perspective, thereby enhancing the predictive power of P-DL.

In the active learning process, one spatial location on the $xy$-plane is initially randomly chosen. Note that, for each selected 2D location, 5 sensors are installed at different depths. Those initial sensor readings will be used to train the P-DL model. After the training is complete, we further apply the AL criterion to determine the next sampling point on the $xy$-plane. We measure the soil moisture at various depths at the new location. The resulting time series data is incorporated into the training dataset to re-train the P-DL model. This active selection iterates itself until the sensor budget is exhausted.
 
\section{Experimental Design and Results}
We validate our P-DAL framework in estimating soil moisture dynamics in both evaporation and infiltration scenarios. Both field geometries are designed as cuboids, configured with 20 nodes in length ($N_L=20$), 20 nodes in width ($N_W=20$), and 10 nodes in depth ($N_D=10$). Both scenarios share the same WRC and HCF constants for a given soil type and condition, with the parameter setting as $K_s=0.0092 \text{ cm/s},~ n = 2,~ m=1.5, ~\alpha=0.0335 \text{ cm}^{-1},~ \theta_s=0.368, ~\theta_r = 0.102$. Note that the distinctions in P-DL modeling for evaporation and infiltration arise from differences in sensor observations and boundary conditions. The groundtruth datasets $\theta(\vect{s},t)$ of the soil system dynamics are obtained from \cite{varela2018implementation}. A Gaussian noise of $\sigma_\epsilon = 0.005$ is introduced to the sensor observation to simulate the measurement noise. Thus, the sensor observation can be represented as $\theta_m(\vect{s},t)= \theta(\vect{s},t)+\epsilon(\vect{s},t)$, where $\epsilon(\vect{s},t) \sim \mathcal{N}(0,\sigma_\epsilon^2)$. 

We assume the total sensor budget is 40. The initial 8 sensing locations are selected on the $xy$-plane, after which 5 sensors are installed at different depths for each of the selected horizontal locations. This sensor placement configuration applies to both active learning and random sampling schemes. Additionally, in order to embed the governing physics into the DNN training, we randomly pick $N_c=10,000$ collocation points from the soil moisture spatiotemporal domain to enforce the RE. The architecture of the neural net is empirically determined to consist of 5 layers, with each layer comprising ten neurons. Model performance is quantified by the relative error ($Er$) defined as:
\begin{equation}
	Er=\frac{\sqrt{\sum_{\vect{s}, t}\|\hat{\theta}(\boldsymbol{s}, t)-\theta(\vect{s}, t)\|^2}}{\sqrt{\sum_{s, t}\|\theta(\boldsymbol{s}, t)\|^2}}
\end{equation}
where $\theta(\vect{s},t)$ and $\hat{\theta}(\vect{s},t)$ stands for the reference and estimated soil moisture levels on the entire spatiotemporal domain, respectively. To evaluate the efficacy of the P-DL model using the training data obtained via the proposed active learning method (termed P-DAL), we undertake a comparative analysis. This analysis compares P-DAL with an alternative approach, where the P-DL model is trained using sensor data derived from non-informative uniform random sampling, named P-DRL.

\begin{figure}
 	\centering
 	\includegraphics[width=3.4in]{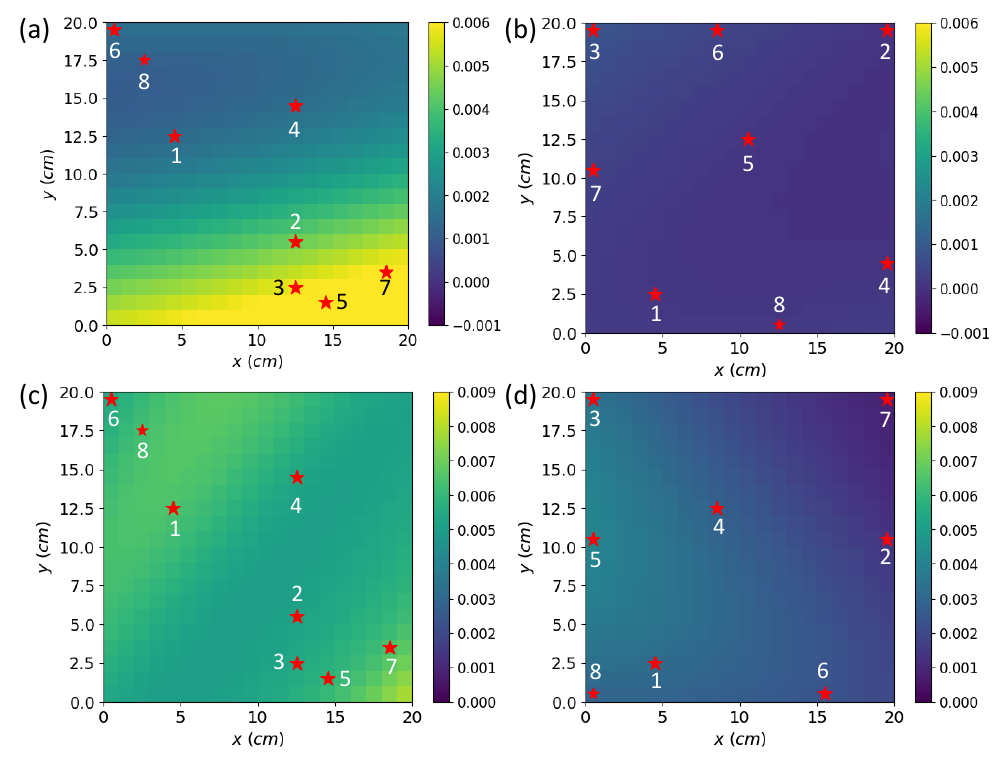}
  \vspace{-20pt}
 	\caption{
  (a-b) The absolute error mapping of $\hat{\theta}$ in the $xy$-plane for the evaporation case, generated by (a) P-DRL and (b) P-DAL. Red stars mark the chosen sensor locations, with numbers indicating the sequence of sensor placement.  (c-d) denote the error mappings for the infiltration case, with sensor placement and selection sequence explicitly marked.
 	}
  \vspace{-15pt}
	\label{Fig:loc_sequence}
 \end{figure}

% \subsection{Evaporation case}
%  \begin{figure*}
% 	\centering
% 	\includegraphics[width=7in]{prec_all.pdf}
% 	\caption{(a) The $Er$ estimated by P-DL model based on volumetric water content sensor data collected from random sampling (P-DRL) and active learning (P-DAL); (b) The evolvement of the volumetric water content across the time domain at an arbitrary node in the 3D cuboid field. (c) The variation of volumetric water content at different depths at an arbitrary location in the $xy$-plane in the infiltration case. 
% 	}
% 	\label{Fig:prec_all}
% \end{figure*}

 \begin{figure*}
 	\centering
 	\includegraphics[width=6.0in]{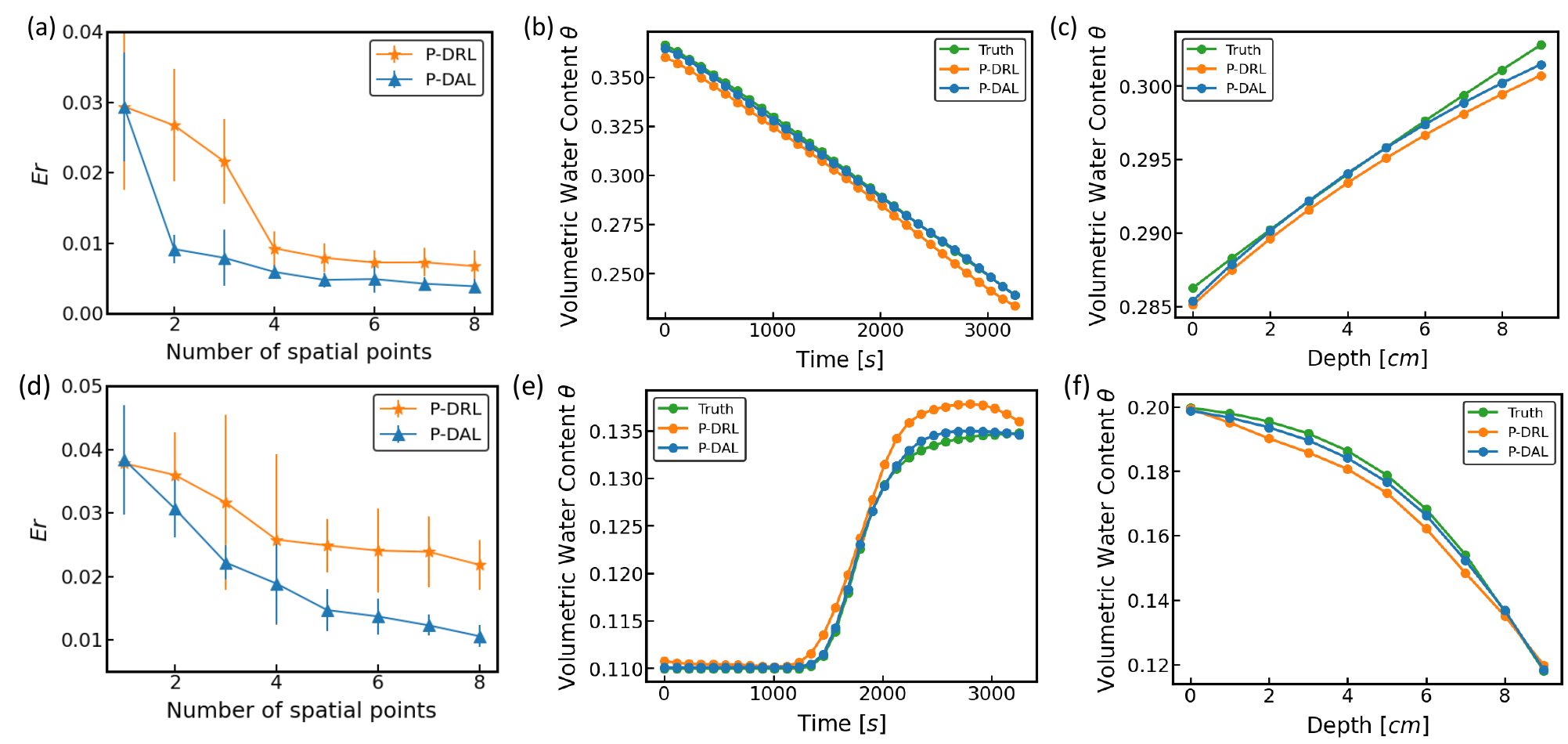}
  \vspace{-8pt}
 	\caption{(a) The relative error $Er$ predicted by the P-DL model using soil moisture sensor data collected from random sampling (P-DRL) and active learning (P-DAL); (b) The evolution of the soil moisture across the time domain at an arbitrary node in the 3D cuboid field. (c) The variation of soil moisture at different depths at an arbitrary location in the $xy$-plane in the evaporation case.
 	}
  \vspace{-8pt}
 	\label{Fig:evap_prec_all}
 \end{figure*}

\subsection{Evaporation Case}
Fig. \ref{Fig:evap_prec_all}(a) illustrates the performance of the P-DL model trained using sensor data gathered via the proposed active learning with uniform random sampling. Specifically, one location on the horizontal plane is selected for each sampling round. For each sampling round, one single location on the horizontal plane is chosen, and 5 soil moisture sensors are uniformly distributed along the vertical axis. To mitigate the variability inherent in network training, the P-DL model is trained 10 times for each sampling round. The mean of the resulting 10 $Er$ values is calculated to ensure consistency in our results. Furthermore, to depict the variability of the $Er$ values, we have included error bars, calculated from the standard deviation, presented in Fig.~\ref{Fig:evap_prec_all}(a). In Fig.~\ref{Fig:evap_prec_all}(b-c), we select the prediction of soil moisture dynamics that exhibits $Er$ closest to the calculated mean. This allows us to showcase a representative model performance that aligns closely with the average prediction accuracy. 

%Fig. \ref{Fig:loc_sequence}(a) presents a groundtruth mapping of volumetric water content in the $xy$-plane at an unspecified depth and time. Figs. \ref{Fig:loc_sequence}(b) and (c) display the predicted mappings by P-DRL and P-DAL, respectively, illustrating the process of selecting optimal sensor placement locations on the $xy$-plane. The comparison in Figs. \ref{Fig:loc_sequence}(b-c) highlights that, compared to the uniform random selection, the active learning approach favored more distributed locations informed with physics residual, enhancing the global accuracy of soil moisture dynamic estimation presented with smaller mapping discrepancy. 

Figs. \ref{Fig:loc_sequence}(a) and (b) show the absolute error in soil moisture between the predictions of P-DRL and P-DAL with the noise-added benchmark, respectively, in the $xy$-plane at an arbitrary depth and time. The figures also illustrate the placement sequence and locations of soil moisture sensors on the $xy$-plane. A brighter color indicates a larger discrepancy from the ground truth. Thus, from Figs. \ref{Fig:loc_sequence}(a) and (b), we see that, unlike uniform random selection, our active learning approach, informed by physics-based residuals, selects more spatially distributed locations. Thus, this strategy improves the global accuracy of soil moisture dynamic estimation, as evidenced by smaller discrepancies (darker color) in the mapping.

As illustrated in Fig. \ref{Fig:evap_prec_all}(a), when the number of spatial points increases, $Er$ reduces. This is because more information on the soil moisture is incorporated into the P-DL model. However, the $Er$ provided by our P-DAL model shows a more rapid decline as opposed to the $Er$ estimated by P-DRL. The $Er$ values given by P-DAL and P-DRL start to separate after the first sampling round. When the number of the selected spatial points on the horizontal plane reaches 8, which means that the sensor budget of 40 is all used, the $Er$ is reduced to $3.89\times 10^{-3}$ by P-DAL, which is 42.4\% less than the $Er$ of P-DRL ($6.75\times 10^{-3}$). This suggests that the proposed P-DAL method can robustly model the soil moisture dynamics, (i.e., volumetric water content), in the crop field with $N = 4000$ spatial nodes at a relative error of  $3.89\times 10^{-3}$ with just 40 sensing locations (8 selected locations on the horizontal plane and 5 sensors installed at different depths for each horizontal location).

Fig.~\ref{Fig:evap_prec_all}(b) presents the evolution of the soil moisture content $\hat{\theta}$ overtime at a specific location, as estimated by the P-DL model. Fig.~\ref{Fig:evap_prec_all}(c) demonstrates the variation of $\hat{\theta}$ along the $z$-axis at a specific 2D location on the horizontal plane at an arbitrary time point. The estimations are based on the data collected from 40 sensors, with the locations of these sensors determined by employing P-DAL or P-DRL. The predictions are benchmarked by the ground truth dynamic evolution (green curve). Both approaches produce good predictions thanks to the physics-based constraint embedded in the P-DL model. However, upon closer examination of Fig.~\ref{Fig:evap_prec_all}(b-c), it becomes evident that P-DAL shows better alignment with the ground truth compared to the P-DRL model. This demonstrates the superior performance of our P-DAL strategy in strategically selecting sensor locations that reduce the variability the most.  %, thereby enhancing the accuracy of modeling spatiotemporal soil moisture dynamics.

%We assume the volumetric water content at 30 discrete time intervals.

\subsection{Infiltration case}
Fig. \ref{Fig:evap_prec_all}(d) illustrates the forecasting ability of the P-DL model when trained with datasets obtained via active learning (i.e., P-DAL) and uniform random selection (i.e., P-DRL). The P-DAL method demonstrates lower $Er$ overall than those produced by the P-DRL. The deviation starts to show up after the first round of sensor placement selection. When the total number of sensors ($N=40$) are all deployed, the $Er$ is reduced to 0.0105 for P-DAL in comparison with the $Er$ of 0.0218 by P-DRL, a 51.8\% difference. %Fig. \ref{Fig:loc_sequence}(d) displays a groundtruth mapping of volumetric water content in the $xy$-plane during precipitation. 
Figs. \ref{Fig:loc_sequence}(b) and (c) show P-DRL and P-DAL predicted mappings, respectively, for the infiltration. Similar to the evaporation case, the active learning strategy optimizes sensor placement for enhanced soil moisture estimation with minimal absolute discrepancy from the ground truth.

Note that the infiltration case presents a more complex soil moisture dynamic pattern compared to evaporation, where the moisture curves tend to be more gradual. This is due to infiltration's heightened sensitivity to external variables, such as rainfall intensity. These factors can cause swift changes in soil moisture levels and create steep moisture gradients as water percolates through the soil. This will increase the difficulty of P-DL model prediction and lead to the estimation error of infiltration higher than the evaporation case. 

Fig. \ref{Fig:evap_prec_all}(e) illustrates the temporal evolution of estimated soil moisture, $\hat{\theta}$, at a designated location, as predicted by the P-DL model. Meanwhile, Fig. \ref{Fig:evap_prec_all}(f) highlights the variation in $\hat{\theta}$ along the vertical direction at a particular 2D point on the horizontal plane, captured at a chosen time point. These estimations are obtained from the output of the DNNs trained by data gathered via the active learning scheme as well as the conventional uniform random sampling method. The prediction accuracy is compared against the actual dynamic development (green curve). Similar to the evaporation scenario, both sampling strategies show an accurate overall trend. However, a detailed review of Figs.~\ref{Fig:evap_prec_all}(e-f) reveals that the P-DAL-generated curves exhibit closer conformity to the empirical data compared to the P-DRL outcomes. This difference reconfirms the effectiveness of our P-DAL approach in identifying optimal sensor placements that would improve the precision of spatiotemporal soil moisture dynamics modeling.

\vspace{-10pt}
\section{Conclusions}
 In this study, we proposed a novel framework for estimating soil moisture dynamics in a 3D cuboid land using noisy soil moisture sensor observations. By embedding the governing physical knowledge and boundary conditions into the DNN framework, this methodology can be extended beyond merely aligning predictions with sensor observations. This allows the predicted soil moisture dynamics to better comply with both the physical principles and sensor observations. Moreover, we develop an innovative active learning methodology to strategically identify a small subset of locations in a large field to deploy soil moisture sensors. This active learning methodology integrates physical residual-based sampling with space-filling design, which provides a more comprehensive, quantitative understanding of soil moisture dynamics. We evaluate the effectiveness of the proposed P-DAL framework in evaporation and infiltration soil scenarios. Results from these numerical experiments show a significant improvement in soil moisture estimation when the active learning methodology is employed to identify the optimal sensor placement compared with random search. %This study holds significant promise for enhancing techniques used in the real-world deployment of soil moisture sensors.

\bibliographystyle{IEEEtran}

\bibliography{references}

\end{document}